# Theoretical knock-outs on biological networks

Pedro J. Miranda* and Sandro E. de S. Pinto

*Department of Physics, State University of Ponta Grossa, Paraná, Brazil*

*Corresponding author: pedrojemiranda@hotmail.com*

and

Murilo S. Baptista

*Institute for Complex System and Mathematical Biology, SUPA, University of Aberdeen, Aberdeen, United Kingdom*

and

Giuliano G. La guardia

*Department of Mathematics and Statistics, State University of Ponta Grossa, Paraná, Brazil*

*Abstract:* In this work we formalize a method to compute the degree of importance of biological agents that participates on the dynamics of a biological phenomenon build upon a complex network. We call this new procedure by theoretical knock-out (KO). To devise this method, we make two approaches: algebraically and algorithmically. In both cases we compute a vector on an asymptotic state, called flux vector. The flux is given by a random walk on a directed graph that represents a biological phenomenon. This vector gives us the information about the relative flux of walkers on a vertex which represents a biological agent. With two vector of this kind, we can calculate the relative mean error between them by averaging over its coefficients. This quantity allows us to assess the degree of importance of each vertex of a complex network that evolves in time and has experimental background. We find out that this procedure can be applied in any sort of biological phenomena in which we can know the role and interrelationships of its agents. These results also provide experimental biologists to predict the order of importance of biological agents on a mounted complex network.

## 1. Introduction

Biological phenomena often rely on a variety of complex dependencies, feed-backs, and auto regulations. As we try to build a general theory to understand these phenomena, one may find it difficult to uncover a common ground to dissertate about biological central questions. One of the most influential theoretical biologists was Nicholas Rachevsky. He introduced the idea that those phenomena could be approached into two different ways: the relational and metrical aspects of biological systems [1].

The relational biology deal with the complexity and relationships between well defined biological agents, such as: enzymes, cells, tissues, organs, species, etc. This sort of approach leads to general structures which can mainly be modeled by graphs, in the terms of graph theory, which is consistent on how reliant on complex structures biological phenomena are. Most of the complex network theory (*i. e.,* graphs that models real entities) developed today for modeling biological systems is a remnant of Rachevsky view point of biological processes. However, there are few works that recognize this author's contribution to this field, and we find a wealth of concepts that can be availed in order to understand biological processes.

On the other hand, there is the metrical biology that encompasses the reducionistic approach of modern biology and biotechnology. This approach allows one to know the biological structures involved on a phenomenon in details, but it rely mostly on the assumption that structure imply function [2]. In other words, the knowledge of all smalls structures involved in a phenomenon implies the knowledge of all its function and leads ultimately to the understanding, description, and prediction of the subject matter.

We recognize that the modern biology brought to light much that we actually know about behavior of biological systems by the knowledge of its parts. However, there still exist difficulties which pervade general realizations of biology itself. Some of the main questions of biology can exemplify these difficulties: what is preserved in life phenomena; how structures are related to give rise to life; how life behaves in time and space, and so on. In order to answer and reconcile these questions it is necessary to build a general theory that turns it possible to approach biology conceptually and experimentally. Modern biology have shown that it has no discourse to offer this theory, since biological phenomena are resilient to final reduction in terms of Cartesian method [2]. This fact of resilience is due to the nature of these phenomena and how we understand them. The lack of a general theory into a metrical approach for biology, pull theoretical biologists towards the relational biology language. One of the most important contribution to this area was the model introduced by Robert

Rosen in 1958 [3]. In his work, Rosen proposed a metabolic network's model that takes into account three metabolycal processes: anabolism, catabolism, and repair. This model is known as $(M, R) - system$, and it is composed by a set of *components* $M_i$ of the system $M$. This system is a connected directed graph in which the vertices are components (*i.e.,* representatives of biological agents) and the directed edges are input and outputs to the components.

The inputs are directed edges that points toward a component and outputs are directed edges that points from components. We interpret inputs and outputs as materials to be utilized by components in order to generate outputs. In this theory, Rosen differentiate coarse structures and fine structure. The latter relates to abstract systems in which vertices are "black boxes", which it is known only its input materials and output materials, but not how it operates; and the former relates to specific known systems of cells, enzymes, tissues, organs – which are results of metrical biology that convey the bridge between the relational biology [3]. After this work, Rosen realized that is wasn't enough to describe most metabolycal phenomena and used this introductory work as a background to posterior formalization: the categorification of $(M, R) - system$ via category theory [4, 5].

A good general theory for a field like biology should allows one to test experimentally, or numerically, each step that the theory when it is developed and detailed. By this assumption we mean that the most propositions should be testable to turn the theory intelligible and scientifically valid. Studying Rosen's model, we find an important proposition that took our attention: the importance of each component of a $(M, R) - system$. Rosen defend that certain components are more important in the operation of the system, and take as a measure of the importance the *number of environmental outputs* of the system that cease to be produced due to a component inhibition [3]. This informal definition of a vertex (*i. e.,* component) importance in a biological systems accounts only on the system outputs to the environment, and has little to do with the internal implication of a vertex inhibition.

If we consider a *non-central component* of a $(M, R) - system$, which is a vertex that if removed would not result in the failure of the entire system, this vertex still should cause more damage than imagined since the mutual reliance of vertex is frequently found in biological systems. For example, the failure of a secondary cytokine can generate debilitation on the complex network in which that cytokine play a role. This is observable because other cytokine, being primary, should compensate the lack of the former inhibited cytokine and the dynamics of the network changes internally, but is generating the same sort and amount of environment outputs. However, when time goes on, the systems changes completely by the inhibition of that secondary cytokine, and some pathology or system malfunction may arise. In another example, lets us consider a cell where exists many enzymes which have the same substrate. The inhibition of them causes a demand in the other, altering the quantity of products generated by the same substrate. As in the case of cytokines, the inhibition of a secondary enzyme causes the primary to be overused, and more transcription of the gene that corresponds that enzyme must be performed. As time progresses, the cellular network will change internally while it is trying to keep its outputs in an acceptable level. A species that have a defined niche, when inhibited in an ecological system, shall cause less competition on species of same niche and limited resources. This inhibition cause variation on how other species are related ultimately changing the structure and dynamical dependences of the complex system it is modeled by. Still, a view from "outside" of the ecological system does not demonstrate great differed before and after removing a species (unless it is a central species), since most environmental outputs are being produced.

Besides subtle, these examples illustrate how deeply is the consequence of removing a system's component, and only the alteration of the system's outputs should not suffice to quantify the degree of importance of a component. In order to remove the subtlety of this conception, this work has as aim generate a steady quantification of the importance of each component involved in a biological network. Inspired on knock-out (KO) on animal models, in which a gene is suppressed and a population of animal subjects is knocked-out for its corresponding phenotype, we introduce the concept of theoretical KO, which is the effect of the removal of a vertex and how it affects the biological network in which it belongs.

We apply this procedure to any biological phenomena that can be modeled by a connected directed graph in which dynamical processes can be used. We proceed this endeavor using random walks in directed graph model, exploring both analytical and computational of this stochastic dynamics. To organize our specific goals to achieve this particular objective we list:

i. propose a generic model in which it is possible to use random walk in directed graphs to define and to calculate an invariant quantity that allows to measure the importance of any vertices of a graph (*i. e.,* complex network);
ii. propose an analytical method to find the same invariant quantity due to the random walk in complex networks, and when applicable;
iii. propose an algorithm that encompasses the sequence of operation over the network to generate statistically the defined invariant quantity;
iv. defend and discuss the experimental utility of the model for predicting important theoretical KOs.

## 2. The network model and the calculation of theoretical KOs

*2.1 Applicable conditions to the model and its justification*

In this section we intend to describe and justify the use of our model, and to orient the interested researcher how to prepare his data to our method. Our starting point is the $(M, R) - system$. As we described above it is a "black block" diagram that is drawn upon a directed connected graph. This means that any two vertices of the system are connected by a directed path. However, for our purposes, this model needs modification in order to generate the quantity that we are interested. For the $(M, R) - system$, the vertices represent biological agents that receive inputs to generated outputs, and is a replaceable component $M_i$ which represents enzymes, organelles, cells, tissues, etc. The directed edges represent materials that are inputs and outputs. Thus, we modify this model by the following conditions:

  *a)* all components are represented by a class of objects that occur in the same place and time and participate in a common biological event, and;
  *b)* the edges are represented by a simply evident interaction, relation, influence or response, between the classes of objects.

The generality of conditions allows our model to be applicable to a wide range of biological known events. By biological event we mean any of the cellular process, such as: citric acid cycle, cellular respiration, DNA transcription, protein folding, and so on. Also, any physiological process, like: hemodynamics, gland functions, neural system responses, pharmacological interactions, immunological behavior, etc. And any ecological system: population relationships, species competition, co-evolution, niche studies, among other cases. In each of these cases, the biologist can identify the set of biological agents that acts and participates in these events: a class of cells of the same type, the class of associated tissues, and the class of individual of a species or the set of species in a community.

The conditions *a)* and *b)* defined above sets the model in the class of phenomenological models, so we can generate phenomenological results that can benefit those who apply it. An initial application of this model, for a very particular case, was realized for the oral tolerance phenomena, in which cellular and humoral components were accounted in a complex network to describe the phenomenon [6]. The results of this particular paper were the calculations of theoretical KOs which was consistent with the experimental KOs in the order of importance for the oral tolerance. In the same manner, we propose the generalization of this model to any biological event that can by approached experimentally and ultimately can test the utility and power of prediction of our model.

The interested researcher can use this model to any of these approaches in any acceptable hierarchical biological level, if the prerequisites are satisfied. This is possible due to the fact that the experimental biologist have difficulty to pinpoint the order of importance of components in a general biological system (*e. g.,* flow charts). Note that all these kind of problems are solved by our model.

In general, our model proportionates a way to identify central components that does not depend only on topological measures in terms of graph theory [7]. However, it is important to capture the relevance of components (*i. e.,* classes of objects) that play a role in both dynamical and topological aspects. As mentioned above, when we study the parts of a complex phenomenon one may lose the track of its interrelationships and its totality. When this happens a strong method should be available in order to preserve the significance of the whole systems which describes the phenomenon of interest.

Clarified all these assumptions and conditions, we formalize the process of creation of a relational network that encompasses a biological phenomenon for the calculation of importance of its components:

  I.    Identify all specific agents that accounts for the phenomenon.
  II.   Consider them as vertices of a directed graph which represent classes of object of the same nature.
  III.  Identify all relationships between the classes of objects.
  IV.   Associate a directed edge to a class of objects that influence other class of object in some way;
  V.    Insert in the graph a vertex that accounts for a *origin* which is regarded, as in Rosen formalism, as the *set of environment inputs*. More specifically, this vertex can only provide directed edges.
  VI.   Insert in the graph a vertex that represents the *terminus*, which is regarded as the *set of environmental output*. More precisely, this vertex only receives directed vertices from any other terminal vertex. A terminal vertex is a vertex that has no directed edges pointing out of it.

This procedure guarantees that the resulting graph is connected and there always exist an *origin* and a *terminus* of the graph [3]. It also guarantees the creation of a biological relational network, which is based on experimental results and observations. This allows one to apply the model that is described in the following section. An example of how a resulting graph constructed by this method is shown in Fig. 1.

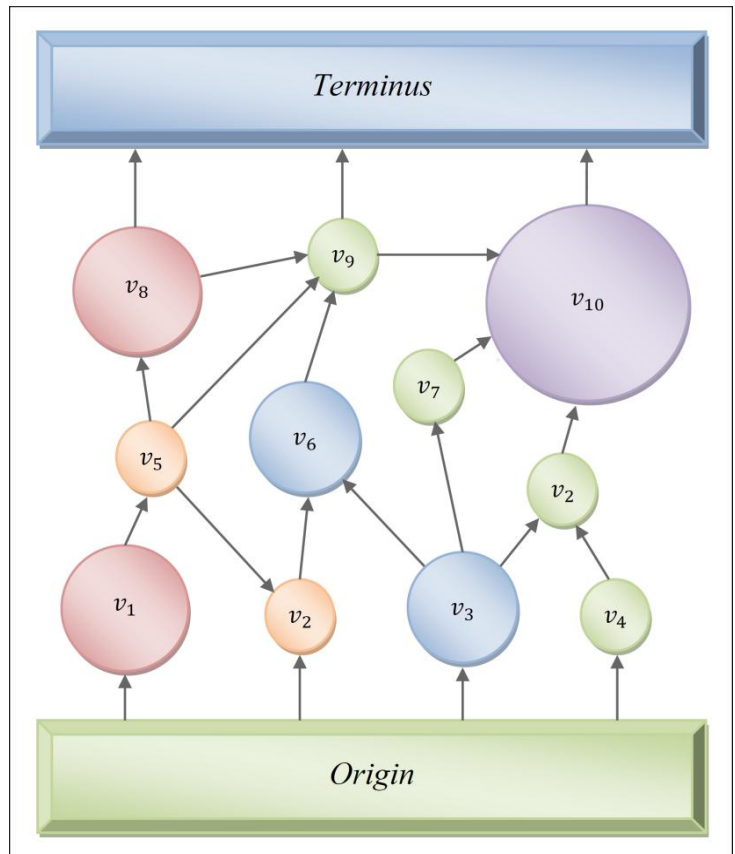

Fig. 1. An example of a relational biological network generated by the process of creation, described in Section 2.1. Where $v_8$, $v_9$ and $v_{10}$ are terminal vertices.

## 2.2 Description of the model

Given a biological relational network constructed as suggested in the former section, we start the study of our model from it. Now let us consider a generic relational network which is fixed (*i. e.,* a network with constant number of vertices and edges). As implied in the $(M, R) - system$, the network is directed and connected. Usually we represent the network by its adjacency matrix: if there is a directed edge connecting the vertex $i$ and $j$, the value of the adjacency matrix's element $a_{ij}$ will be 1, and 0 otherwise. We consider a network formally as a graph $G = (V(G), E(G))$ which is a ordered pair of a set of vertices $V(G) = \{v_1, v_2, ..., v_n\}$ and a set of edges $E(G) = \{e_1, e_2, ..., e_k\}$.

To understand the biological interaction of components in a network, we define the concept of out degree $k_{out}$ for a generic vertex $i$, given by

$$k_{i\,out} = \sum_j a_{ij}, \tag{1}$$

where $a_{ij}$ are the elements of the adjacency matrix. This quantity is expressed in the graph as the number of edges pointing out from the vertex $i$. Establishing these concepts, we are ready to employ the random walk on biological relational networks. Random walks can be understood as a stochastic process in discrete time-steps in which a generic walker follows a path determined by the relational network's topology. Now let us formalize this dynamics into the following conditions:

i. all walkers are created at the *origin* of the network and the total number of walkers on it is given by

$$N(t) = t. \qquad (2)$$

This equation guarantees that in each time step a walker is created on the *origin* and walks one time;

ii. the walker is allowed to shift positions from any vertex $i$ to any other vertex $j$ if and only if $j$ is a first degree neighbor of $i$ (*i. e.*, separated by only one edge);

iii. given the variation of time from $t$ to $t+1$, the probability $p$ of shifting is given by $p = 1/k_{out}$;

iv. the walker must respect the directionality of the edges; if the walker is at vertex $i$, he only can go to other vertex $j$ if and only if there is a edge point from the vertex $i$ to the vertex $j$.

When more walkers are inserted in the network, it is convenient to create a positional walker vector $W(t)$ defined by

$$W_G(t) = \left(w_1(t), w_2(t), w_3(t), \ldots, w_{N(t)}(t)\right), \qquad (3)$$

where $W_G(t)$ is a vector of $t$ coordinates, $w_i(t)$ is the vertex position of the *i-th* walker at time $t$, and $G$ is a directed graph which represents the biological relation network.

Since we are interested in studying the probability of a walker to shift positions, we say that the probability of a generic walker to walk from a vertex $i$ to a vertex $j$ equals the frequency in which any walker at vertex $i$ shifts to $j$. Numerically, this frequency can be estimated once the time that system evolves is large enough. Since the network is directed, the probability of a walker to arrive on the vertex $j$ from the vertex $i$ can be distinct from arriving to $i$ from $j$. Considering this fact, we can compute the concentration of walkers on a component introducing the concept of the position topological vector $S_G(t)$ which is defined by

$$S_G(t) = (\sigma_1(t), \sigma_2(t), \sigma_3(t) \ldots \sigma_n(t)), \qquad (4)$$

where $\sigma_i(t)$ stands for the number of walkers at time $t$ on vertex $i$. These vectors have $n = |V(g)|$ fixed components. As we are interested on the probability of finding a certain frequency of walker to each vertex of a graph, we define the flux of walkers on a vertex $i$ as

$$f_i(t) = \frac{\sigma_i(t)}{N(t)}. \qquad (5)$$

With this formulation, we can build our main state vector of the system at time $t$ of $G$ as

$$F_{V(G)}(t) = \left(f_1(t), f_2(t), f_3(t), \ldots, f_n(t)\right). \qquad (6)$$

We call this vector flux vector, which is also fixed and has $n$ coordinates. Higher values of $f_i(t)$ means that the vertex $i$ is more activated in a relational sense and it can also be understood that more information is passing through this component. This vector is of central interest to this work: for it is a dynamical quantity that depends on the random choice of each walker on the vertices of the network. Since the network is connected, when $t$ tends to infinity the flux vector $F_{V(g)}(t)$ tends to its stationary distribution $F_{V(G)}(t \to \infty)$. We shall later show how to find this stationary distribution in two ways, namely, analytically and by applying an algorithm.

Further, we are concerned on how a topological modification causes dynamical changes. We consider a graph $G'$ a graph derived from $G$ by removing one vertex (*i. e.*, a KO graph). This procedure will be called theoretical KO. Since the topology of the graph is changed, the probabilities of the random walk also changes. This means that we have a new flux vector if the random walk is performed for $G'$, which generates a new stationary distribution $F'_{V(G')}(t \to \infty) = (f'_1(t), f'_2(t), f'_3(t), \ldots, f'_{(n-1)}(t), 0)$, where we assume without lost of generality that the removed vertex was $v_n$. The last component of the former vector is zero due to the impossibility of a walker to be at $v_n$. Since we want to compare the dynamics change due to a KO, we introduce

$$D_{G,G'} = F_{V(G)}(t \to \infty) - F'_{V(G')}(t \to \infty) = (\Delta f_1, \Delta f_2, \Delta f_3, \ldots, \Delta f_n = f_n), \tag{7}$$

where $\Delta f_i = \lim_{t \to \infty}[f_i(t) - f'_i(t)]$, for $1 \le i \le n$ [8]. The values of $\Delta f_i$ can be negative or positive: if it is positive, it means that the activation of the component is increased by the KO; otherwise, it means that the activation is decreased. For each of these cases, we define the relative error by

$$\mu_i = \begin{cases} \dfrac{\Delta f_i}{f_i(t \to \infty)}, & \text{for } \Delta f_i > 0 \\ \dfrac{|\Delta f_i|}{f'_i(t \to \infty)}, & \text{for } \Delta f_i < 0, \end{cases} \tag{8}$$

which scale the variation $\Delta f_i$ with respect to the largest density. We also define the average of the relative error by

$$M_{G,G'} = \frac{\sum_{i=1}^{n} \mu_i}{n}. \tag{9}$$

We call $M_{G,G'}$ by relative mean error. Note that when $M_{G,G'}$ is close to zero, there is no substantial modification in the dynamical behaviour of the network. On the other hand, if $M_{G,G'}$ is close to one this means that the KO vertex plays a fundamental role on the dynamics of the network, ergo in the biological phenomenon in which it participates.

Our quantities rely upon how we understand the biological phenomenon; we propose that the information unchained by the activity of an initial agent (*i. e.*, the component that represents the origin) in the network starts with information processing that influence all other components involved. This information process defines the phenomenological model proposed here. This means that all biological phenomena that can fit into our conditions can be deeply analysed by our method. As mentioned above, our method is able to detect the level of importance of each vertex as well as to describe the dynamics of the biological network of interest. As it is clear from the context, the main goal of our formal model is to compute the values of $F_{V(G)}(t \to \infty)$ and to compare it to $F'_{V(G')}(t \to \infty)$. In the next sections we describe how to find these vectors.

*2.3 Algebraic aspect of the model*

Aiming to seek an analytic way to unravel the values of $F_i(t \to \infty)$, we introduce the transition matrix $T$ of a graph. This matrix is composed by elements which correspond to the probability of a walker to leave the vertex $i$ and arrive at vertex $j$ defined as

$$p_{ij} = \frac{1}{k_{i\,out}}, \tag{10}$$

where $p_{ij}$ are entries of the stochastic matrix $T$. For our interest, we wish to explore the following expression:

$$F_{V(G)}(t+1) = TF_{V(G)}(t), \tag{11}$$

which implies

$$F_{V(G)}(t) = T^t F_{V(G)}(t = 0). \tag{12}$$

Our interest is to compute $F_{V(G)}(t \to \infty)$; thus we have to find $T^t$ when $t$ tends to infinity. Analogically, we can find the vector $F'_{V(G')}(t \to \infty)$ in order to calculate the value of $M_{G,G'}$. In order to assess this mathematical procedure we introduce the non-negative matrices theory [8]. If we could calculate $T^\infty$ by the multiplicating $T$ a infinite number of times, we would note that the entries of the rows of $T^\infty$ is the asymptotic values of the coordinates of $F_{V(G)}(t \to \infty)$. And this vector is the stationary distribution that we can to find for the calculation of KOs. However, since the calculation of $T^\infty$ by iterating the multiplication to the infinity is limited, we will consider that the flux vector is a Markov Chain [8]. This is to say that the transition of states given by the elements of $T$ follows a Markov Chain, so we consider the set $\{\lambda_i\}$ of eigenvalues of $T$. Thus, if the matrix features the Perron-Frobenius we got the following conditions to be satisfied

i. $|\lambda_1| \geq |\lambda_2| \geq |\lambda_3| \ldots \geq |\lambda_n|$,
ii. $|\lambda_1| = 1$, and
iii. $|\lambda_i| \leq 1$ for all $2 \leq i \leq n$.

Given these premises, we operate $T$ on the unity vector **1** and find that $T\mathbf{1} = 1\mathbf{1}$, where 1 is the eigenvalue and **1** a corresponding eigenvector. Thus, since all sums of $T$ are equal, 1 is the Perron-Frobenius eigenvalue of $T$, then we take **1** as the right Perron Frobenius eigenvector. From this, we can calculate the normed vector $\mathbf{v}'$ such that $\mathbf{v}'\mathbf{1} = 1$ (i. e., scalar product) implies that $\mathbf{v}'T = \mathbf{v}'$. This means that the Markoc Chain given by $T$ has a unique stationary distribution $\mathbf{v}$ [8]. For our purposes we have that $\mathbf{v} = F_{V(G)}(t \to \infty)$.

Now we can list the sequence of mathematical procedures in order to find $\mathbf{v}$ for a given graph $G$:

I. Calculate the transition matrix $T$ via equation (10);
II. Test the condition of non-negative matrices of Perron-Frobenius for the set of eigenvalues of $T$;
III. If the matrix $T$ satisfies all conditions, find the eigenvector associated to the eigenvalue that equals 1. If $T$ does not satisfies the conditions, see the next section to find the stationary distribution $F_{V(G)}(t \to \infty)$ algorithmically;
IV. Normalize the eigenvector associated to the eigenvalue 1. The resulting normalized eigenvector is the are stationary distribution $F_{V(G)}(t \to \infty)$.

In a theoretical KO sense, we use this procedure every time we perform modifications on a graph. If an original graph $G$ and its KO associated graph $G'$ follows the Perron-Frobenius criteria, we can compute $M_{G,G'}$ exactly. However, it is not sure that we will always encounter graphs that satisfies these criteria, so we have covered this situation in the following section.

*2.4 Algorithmic aspect of the model*

In this section we begin by describing the details of the algorithm that incorporates all steps presented in Section 2.2. This approach is especially useful in the case that the transition matrix $T$ does not satisfy Perron-Frobenius criteria [8]. The algorithm presented here is implemented in order to generate a mean flux vectors fixing the parameters of a random walk on a directed graph. The parameters of the algorithm are the time $t$ mentioned above, and the number of walks $L$ that is the number of repetitions of walks in a directed graph. Fixing a large enough time, the parameter $L$ is necessary to compute a mean of the flux vector that tends to $F_{V(G)}(t \to \infty)$ (i. e., with respect to the corresponding coordinates) as $L$ tends to infinity. According to this reasoning, we can measure the mean of the flux vector by

$$\bar{F}_G(t, L) = \frac{\sum_{i=1}^{L} [F_{V(G)}(t)]_i}{L}. \tag{13}$$

For each KO performed over an initial graph $G$, we have a mean flux vector $\bar{F}_{G'}(t,L)$ associated to $G'$. In this manner, we call the mean relative error as $\bar{\mu}_i$. Then, for a given KO calculated via this method, we compute the relative mean error by

$$\bar{M}_{G,G'}(t,L) = \frac{\sum_{i=1}^{n} \bar{\mu}_i}{n}. \tag{13}$$

We can provide the implementation of this algorithm in R language [9]; for this, please contact the corresponding author.

## 3. Discussion

### 3.1 General biological implication of the model

Note that both approaches provide values of $M_{G,G'}$ (algebraic) and $\bar{M}_{G,G'}$ (algorithmic). Additionally, we have that $\bar{M}_{G,G'}(t,L)$ tends to $M_{G,G'}$ for a large enough $t$ and $L$. It is preferable to utilize the algebraic method, since it generates the exact quantity of interest. However, this method is only applicable when both transition matrices of $G$ and $G'$ satisfy Perron-Frobenius criteria. We also want to stress out that for each biological agent represent a vertex on a directed graph, we have a value for $M_{G,G'}$. The closer $M_{G,G'}$ is to one, the most important role the vertex plays. This importance lies both in terms of dynamical and topological roles, once the directed graph captures the interrelationship modelling associated to the relational biology [1], and the random walk on it captures the diffusion of how those relationships occur.

A natural question is: why to utilize random walk over other kinds of dynamics? We defend that random walk dynamics is the simplest and powerful way to study diffusion of "stimuli", information, particles, etc [8]. Thus, it is especially useful if one does not have the exactly way to estimate the influences of biological agents over time. More specifically, we proposed this dynamics since it is the simplest possible, powerful, and the most important thing: it is the unique dynamics applicable even in the case that the temporal dependencies of coefficient that dominates the relation of agents are not known. Additionally, our model is of phenomenological nature. This implies that we can only provide phenomenological quantifiers. Besides, this model is an initial stepping-stone to a possible real based model, which can consider dependencies on time of biological agents.

Finally, we want to advise to interested experimental biologists that this model has a powerful of prediction for most biological phenomena that relies upon relational biology. An extensive work has been done for the immunological phenomenon of oral tolerance [6]; however, the viewpoint of the present work is to spread this procedure to other fields of biology. Furthermore, given the generality of the building criteria and the collection of data from the model, our results can be applied in a wide variety of investigations.

### 3.2 Further development of the model

We have built this model to know how information, or "stimuli", on a relational biological network, modelled by a directed graph, behaves in time. As future works, it is interesting to develop this model by considering the degree of "emergence" that a local agent receives from the global dynamics that behaves stochastically. In this manner, we hope that our work can inspire other researchers in this venue.

## 4. Conclusions

The model described in this work is capable of modeling and generating an invariable quantity associated to the random walk on directed network. We have started from a $(M, R) - system$ foundation to turn our approach intelligible. We also have described two ways to encounter the quantifiers generated by the proposed method, which encompasses all possibilities that fits into our build criteria. Finally, our method can predict the order of importance for all biological agents that participate in a complex network modeled by means of a directed graph. Based on these new results, this work is useful for those biologists that are interested on the knowledge of this order. Much of what is done experimentally rely upon this information validating, therefore, the ways implemented along this research.

REFERENCES

1. Rashevsky, N. 1948. *Mathematical Biophysics.* Rev. Ed. Chicago: University of Chicago Press, 1954.
2. Cottam, R., Randon, W., Vounckx, R. 2007. Re-Mapping Robert Rosen's (M, R)-Systems. *Chem. Biod.*, **4**, 2352-2368. doi: 10.1002/cbdv.200790192.
3. Rosen, R. 1958. A relational theory of biological systems. *Bull. Math. Biophysics,* **20**, 245-60.
4. Rosen, R. 1958. The representation of Biological Systems from the Standpoint of the Theory of Categories. *Bull. Math. Biophysics,* **20**, 317-42.
5. Rosen, R. 1959. A relational theory of biological systems II. *Bull. Math. Biophysics,* **20**, 109-28.
6. Miranda, P. J., Delgobo, M., Marino, G. F., Paludo, K. S, da Silva Baptista, M., et al. 2015. The Oral Tolerance as a Complex Network Phenomenon. PLoS ONE, **10**, doi: 10.1371/ journal.pone.0130762
7. Albert, R., Barabási, A-L. 2002. Statisticl mechanics of complex networks. *Rev. of Mod. Phys.,* **74**, 47-94. doi: 10.1103/RevModPhys.74.47.
8. Seneta, E. 1973. *Non-negative Matrices and Markov Chains*. Rev. Printing. New York: Springer Science+Business Media, Inc. 2006.
9. R Core Team. 2012. R: A language and environment for statistical computing. R Foundation for Statistical Computing, Vienna, Austria. ISBN 3-900051-07-0, URL http://www.R-project.org/.